\documentclass[aps,twocolumn,showpacs]{revtex4}
\usepackage{graphicx}
\include{amssym}
\def\PSfig#1#2{\centerline{\scalebox{#1}{\includegraphics{#2}}}}

\begin{document}

\title{OZI violating eight-quark interactions as a thermometer for 
       chiral transitions} 

\author{A. A. Osipov\footnote{On leave from Dzhelepov Laboratory of 
        Nuclear Problems, Joint Institute for Nuclear Research, 
        141980 Dubna, Moscow Region, Russia}, 
        B. Hiller, J. Moreira, A. H. Blin 
        }
\affiliation{Centro de F\'{\i}sica Te\'{o}rica, Departamento de
         F\'{\i}sica da Universidade de Coimbra, 3004-516 Coimbra, 
        Portugal}

\begin{abstract}
This work is a follow-up of our recent observation that in the $SU(3)$ 
f\mbox{}lavor limit with vanishing current quark masses the
temperature for the chiral transition is substantially reduced by
adding eight-quark interactions to the Nambu --- Jona-Lasinio 
Lagrangian with $U_A(1)$ breaking. Here we generalize the case to 
realistic light and strange quark masses and conf\mbox{}irm our prior 
result. Additionally, we demonstrate that depending on the strength of 
OZI violating eight-quark interactions, the system undergoes either a
rapid crossover or a f\mbox{}irst order phase transition. The meson
mass spectra of the low lying pseudoscalars and scalars at $T=0$ are 
not sensitive to the dif\mbox{}ference in the parameter settings that 
correspond to these two alternatives, except for the singlet-octet 
mixing scalar channels, mainly the $\sigma$ meson. 
\end{abstract}

\maketitle



The proposal to include eight-quark interactions to stabilize the
ground state of the combined three f\mbox{}lavor Nambu --- 
Jona-Lasinio (NJL) \cite{Nambu:1961} and 't Hooft \cite{Hooft:1976} 
Lagrangians \cite{Bernard:1988}-\cite{Bernard:1993} has been done in 
\cite{Osipov:2005b}, where the most general spin zero combinations 
have been worked out. The bosonized Lagrangian and the corresponding 
scalar ef\mbox{}fective potential and gap equations have been derived
at leading order of the $1/N_c$ expansion, embracing ef\mbox{}fects of
the dif\mbox{}ferent current quark masses $m_u\ne m_d\ne m_s$.  

Since then a number of physical applications of the enlarged model has 
been discussed at length. Among them the spectra and properties of the 
low lying pseudoscalar and scalar meson nonets \cite{Osipov:2006a},
the behavior of the hadronic vacuum in a constant magnetic
f\mbox{}ield \cite{Osipov:2007a}, or at f\mbox{}inite temperature 
\cite{Osipov:2007b}. This latter study has been done in the $SU(3)$ 
f\mbox{}lavor limit for massless quarks. In particular, it has been 
realized that the critical temperature $T_c$ at which transitions
occur from the dynamically broken chiral phase to the symmetric phase 
decreases in presence of the eight-quark interactions. The part of the 
multi-quark Lagrangian that violates the Okubo-Zweig-Iizuka (OZI) rule 
is mainly responsible for this ef\mbox{}fect. 

Although eight-quark interactions are not required to stabilize the 
vacuum of the $SU(2)$ NJL model, their ef\mbox{}fect on the critical
temperature of the chiral transition has been studied as well for this 
case \cite{Kashiwa:2006}. The conclusion remained unchanged.
 
The OZI violating interactions have a strong impact on the type of 
chiral transition with massless quarks. If their strength is relatively 
weak, the phase transition restoring chiral symmetry at f\mbox{}inite 
temperatures is of second order, as in the standard NJL model. 
However, with increase of the strength at $T=0$ the ef\mbox{}fective 
potential develops a minimum at the origin \cite{Osipov:2005a}. In
this case there is the possibility of coexistence with another minimum 
obtained by spontaneously breaking the symmetry from a certain
critical value of the strength of the 't Hooft interaction. These 
starting conf\mbox{}igurations with double vacua at $T=0$ 
\cite{Bicudo:2006} lead to a lowering of the critical temperature as 
well, but a new interesting feature is that one deals now with a
chiral transition of f\mbox{}irst order. 
 
We stress that the usual picture associated with spontaneous breakdown 
of chiral symmetry in NJL models is the one related with a maximum of 
the potential at the origin, induced by the large value of the 
four-quark coupling strength. The alternative picture that we promote
is only possible due to the $U_A(1)$ breaking terms and eight-quark 
interactions. The latter guarantee global stability of the vacuum and
do not alter the quality of mesonic spectra obtained in the
conventional case.  

The reported decrease in temperature is welcome in view of recent 
lattice calculations \cite{Aoki:2006}, obtained for f\mbox{}inite
values of the quark masses. In this case there is evidence that a
rapid crossover occurs at $T\sim 150\,\mbox{MeV}$ (note that this
value is not reached in the framework of the standard NJL approach). 
In the massless case one expects a f\mbox{}irst order transition for 
the three f\mbox{}lavor case \cite{Wilczek:1984}-\cite{Lenaghan:2000}. 

In this letter we study the combined impact of realistic values for
the current quark masses and eight-quark interactions on the occurence
of extrema in the thermodynamic potential of the model and their
behavior as functions of the temperature. Beside a rapid crossover the 
model also allows for f\mbox{}irst order transitions to occur. The
latter is a new attribute of NJL models, induced by the eight-quark
forces. We elucidate how the strength of eight-quark interactions
versus the four-quark coupling constant gauges the occurence of the 
dif\mbox{}ferent allowed solutions. The temperature dependence of
meson spectra for the low lying pseudoscalars and scalars is 
calculated in the rapid crossover case to illustrate the result. 

We do not consider here the ef\mbox{}fect of the Polyakov loop 
\cite{Polyakov:1978}-\cite{Svietitsky:1986}, which is known to 
increase the transition temperature  by $\sim 25$ MeV
\cite{Aoki:2006}. In relation with the NJL model the role of the 
Polyakov loop has been investigated in several papers, see {\it e.g.} 
\cite{Fukushima:2004}-\cite{Hansen:2007}. The present study can be 
extended likewise. 

The central object of the present analysis are the gap equations at 
f\mbox{}inite temperature. The corresponding expressions at $T=0$ were 
derived in \cite{Osipov:2006a} within a generalized heat kernel scheme 
which takes into account quark mass dif\mbox{}ferences in a symmetry 
preserving way at each order of the long wave-length expansion 
\cite{Osipov:2001}, and we will consider from now on the isospin limit, 
$m_u=m_d\ne m_s$
\begin{equation}
\label{gap}
   \left\{
   \begin{array}{lcr}
   h_u+\displaystyle\frac{N_c}{6\pi^2} M_u
       \left(3I_0-\Delta_{us} I_1 \right)=0, 
   \\
   \\
   h_s+\displaystyle\frac{N_c}{6\pi^2} M_s
       \left(3I_0+2\Delta_{us} I_1 \right)=0.
   \end{array}
   \right.
\end{equation}

This system must be solved selfconsistently with the stationary phase
equations
\begin{equation}
\label{SPA}
   \left\{ \begin{array}{l}
\vspace{0.2cm}   
   Gh_u + \Delta_u +\displaystyle\frac{\kappa}{16}\, h_uh_s
   +\displaystyle\frac{g_1}{4}\, h_u(2 h_u^2+h_s^2)
   +\displaystyle\frac{g_2}{2}\, h_u^3=0, \\
\vspace{0.2cm}   
   Gh_s + \Delta_s +\displaystyle\frac{\kappa}{16}\, h_u^2
   +\displaystyle\frac{g_1}{4}\, h_s(2 h_u^2+h_s^2)
   +\displaystyle\frac{g_2}{2}\, h_s^3=0. 
   \end{array} \right.
\end{equation}
Here $\Delta_{us}=M_u^2-M_s^2$, $\Delta_l=M_l-m_l$, $l=u,d,s$, and
$M_l$ denote the constituent quark masses. The factors $I_i$, are 
given by the average $I_i = [2J_i(M_u^2)+J_i(M_s^2)]/3,$
$i=0,1,\ldots$, and represent one-quark-loop integrals 
\begin{equation}
\label{ji}
   J_i(M^2)=\int\limits_0^\infty\frac{{\rm d}t}{t^{2-i}}\rho 
   (t\Lambda^2) e^{-t M^2},
\end{equation}
with the Pauli-Villars regularization kernel \cite{Osipov:2004}
\begin{equation}
   \rho (t\Lambda^2)=1-(1+t\Lambda^2)\exp (-t \Lambda^2),
\end{equation}
where $\Lambda$ is an ultra-violet cutof\mbox{}f (the model is not
renormalizable). In the approximation considered we need only to know
\begin{equation}
\label{j0}
   J_0(M^2)=\Lambda^2- M^2\ln\left(1+\frac{\Lambda^2}{M^2}\right),
\end{equation}
and
\begin{equation}
\label{j1}
      J_1(M^2)=\ln\left(1+\frac{\Lambda^2}{M^2}\right)  
      -\frac{\Lambda^2}{\Lambda^2+M^2}\ .
\end{equation}

The model parameters are the four-quark coupling $G$, the 't Hooft 
interaction coupling $\kappa$, the  eight-quark couplings $g_1,g_2$ 
($g_1$ multiplies the OZI violating combination), the current quark
masses $m_l$, and the cutof\mbox{}f $\Lambda$. The stability of the 
ef\mbox{}fective potential is guaranteed if the couplings
fulf\mbox{}ill the following inequality \cite{Osipov:2005b}
\begin{equation}
\label{ineq1}
   g_1>0, \quad g_1 +3g_2>0, \quad 
   G>\frac{1}{g_1}\left(\frac{\kappa}{16}\right)^2.
\end{equation}  

We need additional input to fix parameters. The mass formulae of the 
pseudoscalar and scalar mesons, and the weak decay couplings $f_\pi,
f_K$ obtained in \cite{Osipov:2006a} are used for that. The f\mbox{}it 
shows the interesting correlation between $G$ and the OZI violating 
coupling $g_1$: $G$ goes down, when $g_1$ increases. The mass of 
$f_0(600)$ is the main observable responsive to these changes, 
diminishing with increasing strength of the $g_1$ coupling. The point
is that although spectra are essentially not sensitive to this     
correlation, the ef\mbox{}fective potential is. We will see that 
f\mbox{}inally the pattern of the f\mbox{}inite temperature
restoration of chiral symmetry depends on how strong the OZI violating 
forces are. 
   

In Tables 1-3 the low lying pseudoscalar and scalar meson 
characterisitics are displayed at $T=0$, (stars denote input). One
sees from Table 1 how a smaller value of the parameter $G$ is 
accompanied with an increase of the eight-quark coupling $g_1$.

\vspace{0.5cm}
\noindent
{\small Table 1 \\
\noindent Parameters of the model: $m_u=m_d, m_s$, and $\Lambda$ are
given in MeV. The couplings have the following units: $G$
(GeV$^{-2}$), $\kappa$ (GeV$^{-5}$), $g_1,\, g_2$ (GeV$^{-8}$). We 
also show the values of constituent quark masses $M_u=M_d$ and $M_s$ 
in MeV (only the case of global minima).}
\vspace{0.1cm}

\noindent
\begin{tabular}{lrrrrrrrrr}
\hline
Sets &$m_u$  &$\ m_s$  &$\ M_u$  &$\ M_s$  &$\ \Lambda$  &$\ G$  &$-\kappa$    
     &$g_1$  &$\ g_2$  
\\ 
\hline
a  & 5.8  & 183 & 348 & 544 & 864  & 10.8  & 921   & 0*      &0*   \\ 
b  & 5.8  & 181 & 345 & 539 & 867  & 9.19  & 902   & 3000*   &-902 \\ 
c  & 5.9  & 186 & 359 & 544 & 851  & 7.03  & 1001  & 8000*   &-47  \\ 
d  & 5.8  & 181 & 345 & 539 & 867  & 5.00  & 902   & 10000*  &-902 \\ 
\hline
\end{tabular}   
\vspace{0.2cm} 

\vspace{0.3cm}
\noindent
{\small Table 2 \\
\noindent The masses, weak decay constants of light pseudoscalars 
(in MeV), the singlet-octet mixing angle $\theta_p$ (in degrees), 
and the quark condensates $\big <\bar uu\big >, \big <\bar ss\big >$ 
expressed as usual by positive combinations in MeV.}
\vspace{0.1cm}

\noindent
\begin{tabular}{lrrrrrrrrr}
\hline
   &$m_\pi$ &$m_K$  &$m_\eta$ &$m_{\eta'}$ &$f_\pi$ &$f_K$ 
   &$\theta_p $  &$-\big <\bar uu\big >^{\frac{1}{3}}$ 
   &$-\big <\bar ss\big >^{\frac{1}{3}}$ \\ 
\hline
a  &138*  &494*  &480  &958*  &92*  &118*  & -13.6  &237  &191  \\ 
b  &138*  &494*  &480  &958*  &92*  &118*  &-13.6   &237  &192  \\ 
c  &138*  &494*  &477  &958*  &92*  &117*  & -14.0  &235  &187  \\ 
d  &138*  &494*  &480  &958*  &92*  &118*  & -13.6  &237  &192  \\
\hline
\end{tabular}
\vspace{0.2cm}

\vspace{0.2cm}
\noindent
{\small Table 3 \\
\noindent The masses of the scalar nonet (in MeV), and the 
corresponding singlet-octet mixing angle $\theta_s$ (in degrees).}
\vspace{0.1cm}

\noindent
\begin{tabular}{lrrrrr}
\hline
   Sets &$\ m_{a_0 (980)}$  &$\ m_{K_0^* (800)}$  &$\ m_{f_0(600)}$ 
        &$\ m_{f_0(980)}$  &$\ \ \ \theta_s$  \\ 
\hline
a    &963.5   &1181   &707   &1353   &24  \\
b    &1024*   &1232   &605   &1378   &20  \\ 
c    &980*    &1201   &463   &1350   &24  \\
d    &1024*   &1232   &353   &1363   &16  \\
\hline
\end{tabular}
\vspace{0.5cm}

The generalization to f\mbox{}inite temperature of these expressions 
occurs in the quark loop integrals $J_0,J_1$. 
Introducing the Matsubara frequencies 
\begin{eqnarray}
\label{j0t}
   &&J_0(M^2)\rightarrow J_0(M^2,T) 
      =16\pi^2 T\!\sum_{n=-\infty}^{\infty} 
      \int\!\frac{{\rm d}^3{p}}{(2\pi )^3} \nonumber \\
   &&\times\!\int\limits_0^\infty\!{\rm d}s\, \rho (s\Lambda^2) 
      e^{-s[(2n+1)^2\pi^2 T^2+\vec{p}^2+M^2]},
\end{eqnarray}
and using the Poisson formula
\begin{equation}
   \sum_{n=-\infty}^{\infty} F(n) = \sum_{m=-\infty}^{\infty}
   \int_{-\infty}^{+\infty} {\rm d}x\, F(x) e^{i2\pi mx},
\end{equation}
where $F(n)=\exp [-s(2n+1)^2\pi^2 T^2]$, one integrates over the 
3-momentum $\vec{p}\,$, getting the result
\begin{eqnarray}
\label{j0t2}
   J_0(M^2,T)
   &\!\!=\!\!&\int\limits_0^\infty
             \frac{{\rm d}s}{s^2}\,\rho (s\Lambda^2) 
             e^{-s M^2}  \nonumber \\
   &\!\!\times\!\!&\left[ 1+2\sum_{n=1}^{\infty} (-1)^{n}
              \exp\left(\frac{-n^2}{4sT^2}\right)\right].
\end{eqnarray}
Similarly one gets
\begin{equation}
\label{j1t2}
    J_1 (M^2,T) = -\frac{\partial}{\partial M^2}\, J_0 (M^2,T). 
\end{equation}
It is then easy to verify that $\lim_{T\to\infty}J_{0,1}(M^2,T)=0,$ and
at $T=0$ one recovers the starting expressions (\ref{ji}). 
  
Using these expressions in Eq. (\ref{gap}), we solve the system 
(\ref{gap})-(\ref{SPA}) numerically. We henceforth assume that the model 
parameters $G, \kappa , g_1, g_2, m_l, \Lambda$ do not depend on 
the temperature. As a result we obtain the temperature dependent 
solutions $M_l(T)$, representing the extrema of the thermodynamic 
potential. 
   
\begin{figure}[t]
\PSfig{0.33}{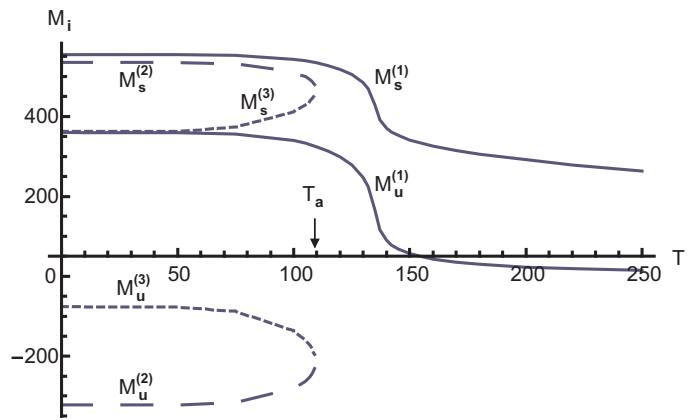}
\caption{Branches of $M_u^{(i)}(T), M_s^{(i)}(T)$ pairs, denoting 
         extremal points of the thermodynamic potential. 
         Solid lines (start at $T=0$ as deepest minima), dashed lines 
         (start as relative minima at $T=0$) and dotted lines (saddle 
         at $T=0$) for the parameter set (c). Only one branch is in
         the physical (positive masses) region (solid curves). All
         units are in MeV.} 
\label{fig1}
\end{figure}
\begin{figure}[h]
\PSfig{0.33}{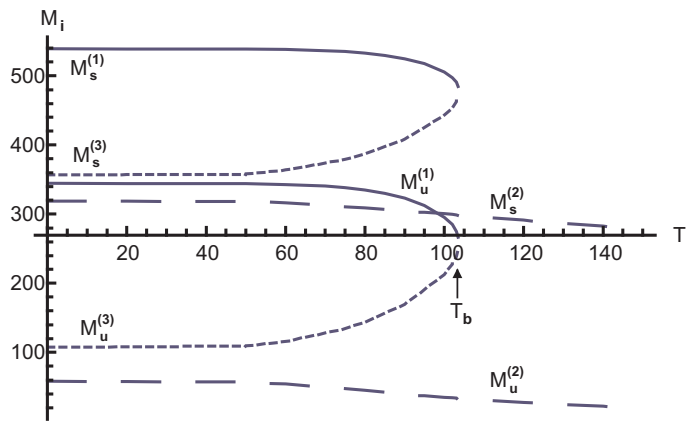}
\caption{The same as Fig. 1 but for set (d).  All branches lie
         in the physical region and coexist up to $T=T_b$, from this 
         value on only one branch survives, with much lower values of 
         $M_l$.} 
\label{fig2}
\end{figure}

The result is show in Figs. \ref{fig1} (set c) and \ref{fig2} (set
d). There are either one or three $(M_u^{(i)} = M_d^{(i)},
M_s^{(i)})$, $i=1,2,3$ couples of solutions at f\mbox{}ixed values of 
$T$. For set (c) (as well as (b)) only one branch of solutions is 
physical, {\it i.e.} positive valued. The other two have negative
values for the light quark masses. This is in contrast with the
$SU(3)$ limit case with zero current quark masses, where one branch 
collapses to the origin $M_u=M_d=M_s=0$ for all values of the 
remaining model parameters and $T$. 

For f\mbox{}inite values of the current quark masses the corresponding 
branch moves in the $M_u, M_s$ plane: only below a certain critical 
value of $G$ does it remain positive valued at all $T$. This is the 
case shown in Fig. \ref{fig2} for set (d), where all three branches 
are positive valued. 


In case (c) the physical changes in the values of the quark masses as 
functions of $T$ are traced back to a single branch. One sees however 
that the onset of the transition occurs at a value of $T=T_a$ for
which the other unphysical two branches meet and cease to exist. The 
rapid crossover occurs in the short temperature interval 
$125\ \mbox{MeV}<T<140 \ \mbox{MeV}$.  

As opposed to this scenario all three branches of set (d) are positive 
valued. Two of the branches (starting from the stable minimum and the 
saddle solution at $T=0$) merge in the physical region at a certain
$T_b$ and the surviving branch has a signif\mbox{}icantly lower mass 
value. Therefore in case (d) the changes involve a jump from one
branch to the other, and lead to discontinuities in the observables. 
This is a f\mbox{}irst order phase transition.   

At present lattice QCD data have not unambiguously settled the
question about the order of the chiral transition. For physical values 
of the quark masses, calculations with staggered fermions favor a smooth 
crossover transition \cite{Brown:1990}, while calculations with Wilson 
fermions predict the transition to be f\mbox{}irst order 
\cite{Iwasaki:1996}. 

Our calculations show that OZI violating interactions can be important 
for the issue. If they are small (as in set (c)) the chiral symmetry 
restoring transition is crossover. However, above some critical value 
of the eight-quark coupling $g_1$ (as in set (d)) the pattern of the 
transition is switched to f\mbox{}irst order. This correlation can be 
used to set an upper (or lower) bound for the strength of the OZI 
violating coupling $g_1$. Without eight-quark interactions (set (a)) 
the physical branch evolves as function of $T$ qualitatively as in
Fig. \ref{fig1}, but the crossover takes place at much larger 
temperatures, $T\simeq 210$ MeV, and the transition is much 
smoother. 

\begin{figure}[t]
\PSfig{0.43}{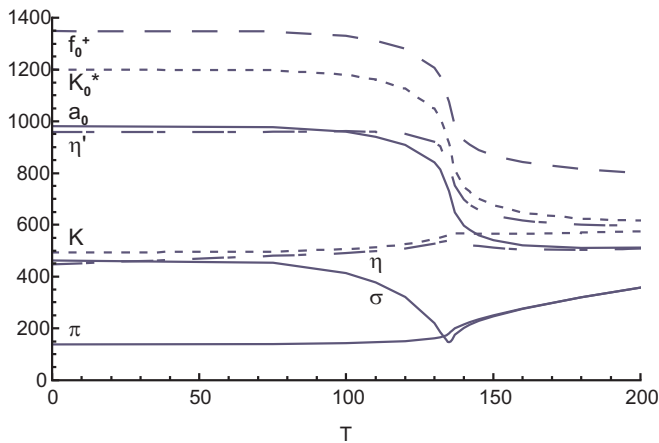}
\caption{The masses of pion, $\sigma = f_0(600)$, $\eta$, kaon, 
         $\eta'$, $a_0$, $K^*_0$ and $f_0^+ = f_0(980)$ from bottom to top, 
         for set (c) as functions of the temperature (all in MeV).} 
\label{fig3}
\end{figure}

The masses of scalar and pseudoscalar mesons at f\mbox{}inite 
temperature obtained for the set (c) are shown in Fig. \ref{fig3}. As 
can be seen there is a rapid crossover for all meson masses in the
same temperature interval as in Fig. \ref{fig1}. Strictly speaking, 
neither this rapid crossover nor the f\mbox{}irst order transition
case (d) do imply restoration of chiral or $U_A(1)$ symmetry, but only 
the recovery of a distorted Wigner Weyl phase, with the minimum of the 
thermodynamic potential shifted to f\mbox{}inite quark mass values due 
solely to f\mbox{}lavor breaking ef\mbox{}fects. 

To get an insight in the role played by the dif\mbox{}ferent
multi-quark interactions we analyze two limits, with the parameter set 
(c) as starting condition. 

Case 1: We set $g_1=g_2=\kappa=0$ and remaining parameters as in (c). 
In this limit the gap equation has only one solution for the
considered parameter set, thus the system is in a distorted 
Wigner-Weyl phase. 

Case 2: We set $\kappa=0$ and all other parameters fixed as in (c). 
In this case there is no $U_A(1)$ breaking, but OZI violating effects 
are present. We verify that in this limit the gap equation has also
only one solution, being again in a distorted Wigner-Weyl phase. 

Thus the spontaneous symmetry breakdown seen in the full set (c) at 
$T=0$ (and also in sets b and d) is driven exclusively by the 't Hooft 
interaction strength $\kappa$. We wish not to include case (a) in the 
present discussion, as it violates the stability conditions of 
(\ref{ineq1}). 

In conclusion, the present study reveals that eight-quark
interactions, whose ef\mbox{}fects are in an almost "dormant" state as
far as the low lying scalar and pseudoscalar mesonic spectra are 
concerned, are of great importance in the study of temperature 
ef\mbox{}fects in chiral multi-quark interaction lagrangians. It turns
out that the mesonic spectra built on the spontaneously broken vacuum 
induced by the 't Hooft interaction strength, as opposed to the
commonly considered case driven by the four-quark coupling, undergo a 
rapid crossover to the unbroken phase, with a slope and at a
temperature which is regulated by the strength of the OZI violating 
eight-quark interactions. This strength can be adjusted in consonance 
with the four-quark coupling (keeping the remaining model parameters
f\mbox{}ixed), and leaves the spectra unchanged, except for the sigma 
meson mass, which decreases. This ef\mbox{}fect also explains why in
the crossover region the sigma meson mass drops slightly below the
pion mass. A f\mbox{}irst order transition behavior 
is also a possible solution within the present approach. Additional 
information from lattice calculations and phenomenology is necessary 
to f\mbox{}ix f\mbox{}inally the strength of interactions. We expect
that the role of eight-quark interactions are of equal importance in 
studies involving a dense medium and extensions of the model with the 
Polyakov loop.  
     
\vspace{0.5cm}
{\bf Acknowledgements}
This work has been supported in part by grants provided by 
Funda\c c\~ao para a Ci\^encia e a Tecnologia, POCI/FP/63412/2005, 
POCI/FP/63930/2005 and POCI/FP/81926/2007. This research is part of 
the EU integrated infrastructure initiative Hadron 
Physics project under contract No.RII3-CT-2004-506078. 


\end{document}